\documentclass[
reprint,
 amsmath,amssymb,
 aps,
floatfix,
]{revtex4-2}

\usepackage{graphicx}% Include figure files
\usepackage{subfigure}
\usepackage{dcolumn}% Align table columns on decimal point
\usepackage{bm}% bold math
\usepackage{float}
\usepackage{upgreek}
\usepackage{lineno}
\usepackage[none]{hyphenat}
\usepackage{soul}
\usepackage{xcolor}
\usepackage{color}
\usepackage{amsmath}
\makeatletter
\def\url@leostyle{
\gdef\UrlFont{\rmfamily}
}
\makeatother
\usepackage{hyperref}
\hypersetup{
  colorlinks=true,  
  linkcolor=blue,  
  citecolor=blue, 
  urlcolor=blue    
}
\begin{document}
%\linenumbers
\preprint{APS/123-QED}

\title{Orbit-Controlled Generation of Two-color Attosecond Mode-locked Free-electron Lasers}% Force line breaks with \\
\author{Lingjun Tu\textsuperscript{1}}
\author{Hao Sun\textsuperscript{1}}%
\author{Huaiqian Yi\textsuperscript{1}}%
\author{Li Zeng\textsuperscript{1}}%
\author{Yifan Liang\textsuperscript{1}}%
\author{Yong Yu\textsuperscript{1}}
\author{Xiaofan Wang\textsuperscript{1}}
\email{wangxf@mail.iasf.ac.cn}
\author{Weiqing Zhang\textsuperscript{2}}
\email{weiqingzhang@dicp.ac.cn}
\affiliation{
		\textsuperscript{1}Institute of Advanced Light Source Facilities, Shenzhen 518107, China\\
        \textsuperscript{2}State Key Laboratory of Chemical Reaction Dynamics, Dalian Institute of Chemical Physics, Chinese Academy of Sciences, Dalian 116023, China}
        
\date{\today}% It is always \today, today,

\begin{abstract}

The generation of attosecond X-ray pulses has garnered significant attention within the X-ray free-electron laser (FEL) community due to their potential for ultrafast time-resolved studies. Such pulses enable the investigation of electron dynamics with unprecedented temporal resolution, opening new avenues in fields such as quantum control and ultrafast spectroscopy. In an FEL, the mode-locking technique synthesizes a comb of longitudinal modes by applying spatiotemporal shifts between the co-propagating radiation and the electron bunch. Here, we propose a novel scheme for generating two-color attosecond mode-locked FEL pulses via orbit control in two-stage mode-locked undulators. Specifically, a chicane with a wiggler inserted in the middle generates periodic temporal-transverse modulation in the electron beam. In this configuration, the low-energy and high-energy components of the beam lase in separate undulator sections, with each producing attosecond mode-locked FEL pulses. Three-dimensional simulations with realistic parameters confirm that trains of 250-attosecond soft X-ray pulses at the gigawatt level can be independently generated in each undulator section. Furthermore, time delay between the two pulse trains can be adjusted over a range of several hundred femtoseconds.

\end{abstract}

\maketitle

\section{INTRODUCTION}
Attosecond pulses are critical for investigating ultrafast phenomena \cite{pupeza2021}.
First generated using high-order harmonic generation (HHG) \cite{hentschel2001}, these pulses primarily occupy the extreme-ultraviolet (XUV) and soft X-ray regions (up to several hundred eV) \cite{kuhn2017eli}. % Over recent decades, 
Free-electron lasers (FELs) \cite{mcneil2010,emma2010} have overcome the intensity and photon energy limitations of HHG-based radiation sources, providing an alternative path to high-brightness, ultrashort radiation pulses. %  \cite{franz2024,yan2024}.
Over the past decade, attosecond science within the FEL community has advanced rapidly, driven by a range of novel schemes including nonlinear bunch compression \cite{huang2017,yan2024}, chirped-pulse compression \cite{wang2025}, self-modulation \cite{duris2020} and cathode shaping combined with superradiance \cite{franz2024}.
Mode-locking FEL schemes \cite{thompson08,kur2011,feng2012,xiang2012,Dunning13} generate a train of ultrashort X-ray FEL pulses with equally spaced, phase-locked modes. In the nominal mode-locking FEL scheme \cite{thompson08}, the electron beam is periodically modulated by an external multi-cycle laser, imprinting a structured energy profile that serves as a prerequisite for mode-locking. Subsequently, a series of chicane magnets positioned between undulators precisely control the radiation delay, ensuring synchronization between the modulation period and FEL resonance. 
In recent years, ultrashort X-ray FEL pulse trains have been experimentally generated  \cite{duris2021,prat2023,hu2025}, enabling the extension of techniques established at lower photon energies, such as Ramsey-comb spectroscopy \cite{morgenweg14}, time-domain ptychography \cite{spangenberg15} 
and quantum optics \cite{reiche2022}. 

Two-color attosecond mode-locked FELs represent a natural extension of single-color pulse-train generation and are of particular interest for future X-ray applications,  including dual-comb spectroscopy \cite{coddington16,suh2016,yu2018}.
Generally, two-color FEL pulses are generated either by manipulating electron beam properties \cite{lutman2014,malyzhenkov20} or by tuning undulator resonance \cite{prat2022,guo2024}. 
Over the last decade, with sufficiently exquisite control of the electron beam orbit, the fresh-slice multicolor method \cite{lutman16,guetg18,wang24} has been demonstrated as an effective approach for generating fully saturated, two-color or multicolor FEL pulses. This method tailors the electron bunch with a temporal-transverse correlation and fine-tunes the electron orbits in the undulator. Such a correlation can be introduced via the strong transverse wakefield of a dechirper \cite{lutman16} or by introducing a dispersive section downstream of the linac \cite{guetg18}. However, the range of the induced temporal-transverse correlation is comparable to the electron bunch length, which limits the minimum generated FEL pulse duration to several femtoseconds. 

\begin{figure*}[ht]
    \centering
    \includegraphics[width=1\linewidth]{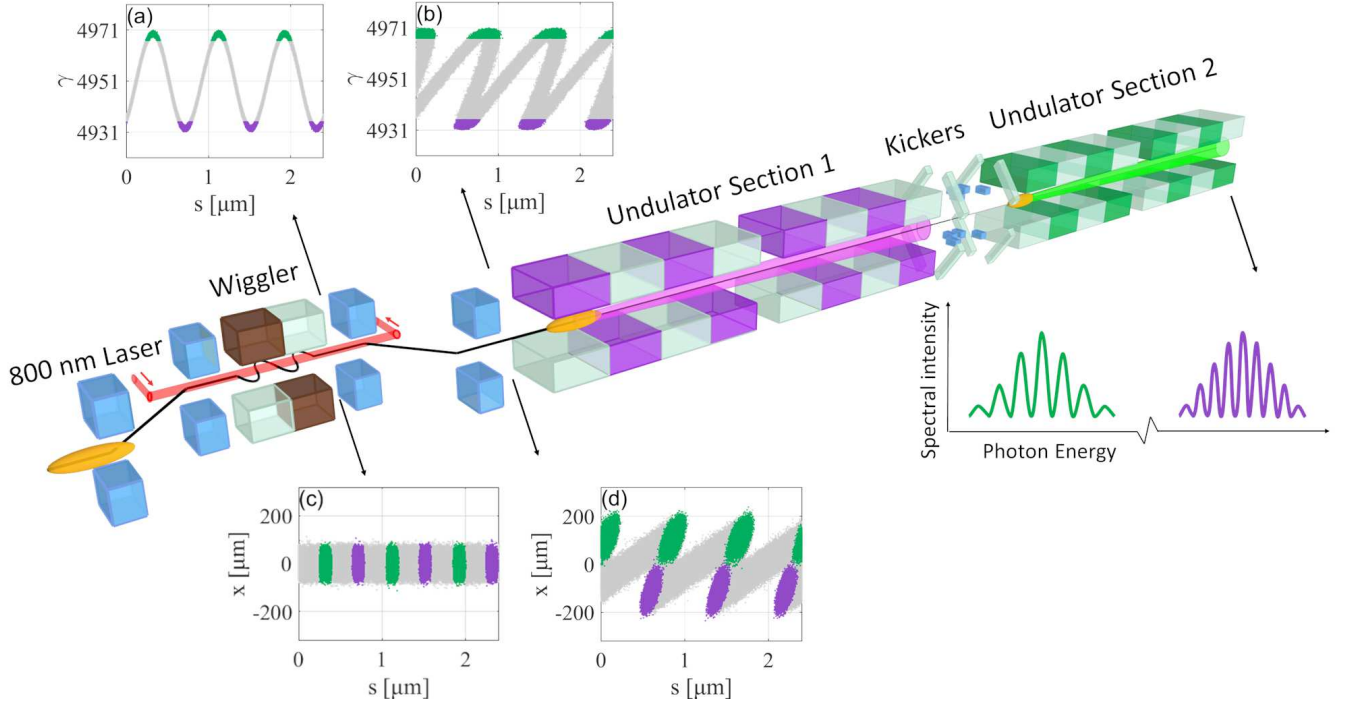}
    \caption{Schematic of the proposed scheme. 
(a) longitudinal phase space and (c) transverse distribution at the end of the wiggler; (b) longitudinal phase space and (d) transverse distribution at the end of the downstream dogleg.
}
    \label{fig1}
\end{figure*}
Here, we propose a novel method for generating two-color attosecond mode-locked XFEL pulse train pairs.
In the proposed scheme, a wiggler is embedded between two doglegs of a magnetic chicane, enabling periodic energy modulation of the electron beam via interaction with an external laser. In the downstream dogleg, low-energy electrons and high-energy electrons are transversely separated, effectively converting the energy modulation into a structured temporal-transverse modulation. 
The modulated beam then enters a two-stage undulator system, with each section comprising short undulators and small chicanes satisfying the mode-locking condition \cite{thompson08}. In the first stage, low-energy electron slices remain on-axis and generate the first-color mode-locked FEL. Kickers then steer the high-energy electrons back to the beam center in the second stage, where they lase to produce the second-color FEL. The central wavelengths of the two pulse trains are continuously and independently tunable by adjusting the resonance conditions in each undulator section.
Within a single modulation cycle, the on-axis electrons occupy only a fraction of the  2.67 fs optical period of an 800 nm laser. Consequently, the individual pulses within the generated two-color mode-locked FEL reach durations of a few hundred attoseconds. 

\section{Principle and Physical Implementation} % 

\begin{table}[thb]
\caption{\label{tab:table1}
Parameters of the electron beam and the equipment used in the simulations.
}
\begin{ruledtabular}
\begin{tabular}{lcdr}
\textrm{Parameter}&
\textrm{Value}\\
\colrule
\textbf{Electron beam}\\
Initial energy $E$ &  2.53 GeV\\
Initial current $I$ &  3 kA\\
Emittance $\epsilon$ &  0.4 mm mrad\\
Uncorrelated energy spread $\sigma_E$ &  180 keV\\
\colrule
\textbf{External laser}\\
Wavelength  $\lambda_s$ & 800 nm\\
Peak power & 40 GW\\
Rms pulse duration & 100 fs\\
\colrule
\textbf{Wiggler}\\
Period & 15 cm\\
Period number & 6\\
$K_w$ & 41.89\\
\colrule
\textbf{Undulator sections}\\
Period & 3 cm\\
Period number per module& 20\\
FODO length & 2 m\\
$K_1$/$K_2$ & 2.118/2.807 \\
FEL wavelength in section 1/2 &  2/3 nm\\
Delay in section 1/2 &  760/740 nm \\
\end{tabular}
\end{ruledtabular}
\end{table}

The layout of the proposed scheme is shown in Fig.~\ref{fig1}.
A periodic energy modulation of the electron beam is induced through interaction with an external laser in a wiggler. 
The downstream dogleg converts the energy modulation into a structured temporal-transverse correlation, while the upstream dogleg maintains beam alignment and mitigates unwanted dispersion.   
Figure~\ref{fig1}(a) shows a representative longitudinal phase space distribution, where low-energy and high-energy electrons are highlighted in purple and green, respectively.
The gray part represents electrons with large energy chirp, where the FEL gain is suppressed \cite{saldin2006,thompson08}. 
Figure~\ref{fig1}(c) shows the initial transverse distribution of the electron beam, while Fig.~\ref{fig1}(b) and Fig.~\ref{fig1}(d) show the energy and transverse distributions at the entrance of the first section, respectively. 
These subfigures are depicted from simulation using typical parameters of Shenzhen Superconducting Soft X-Ray Free-electron Laser (S$^3$FEL) \cite{wang2023}, as listed in Table~\ref{tab:table1}. 
The electron beam extracted from a linear accelerator was characterized by the following parameters: an energy of 2.53 GeV, a flat current of 3 kA, rms normalized emittance of 0.4 mm mrad, and an uncorrelated energy spread of 180 keV. 
The interaction between the electron beam and the external laser occurs in a six-period wiggler with a total length of 0.9 m. An external laser with a wavelength of 800 nm, peak power of 40 GW, beam waist of 0.5 mm, and rms pulse duration of 100 fs is employed, all of which are attainable using state-of-the-art laser systems \cite{petrov2007, tsai2022}. The resulting energy modulation amplitude is approximately 10 MeV. 
In the simulation of Fig. 1, the dispersion and the momentum compression factors of each dogleg are respectively 30 mm and 90 $\upmu $m to illustrate the basic principle. At the exit of the downstream dogleg, the transverse separation between the highest- and lowest-energy electrons reaches approximately 200 $\upmu $m.

The wavelength of the first color is selected to be a typical soft x-ray wavelength of 2 nm. The second color is chosen to be 3 nm, to differentiate the proposed scheme from those utilizing microbunching at harmonics \cite{guo2024}.
For the two undulator sections,
the first satisfies the conditions for mode locking of the low-energy electron slices, while the second satisfies the conditions for mode locking of the high-energy electron slices. 
Specifically, the sum of the slippage in a single undulator module and the slippage in the following chicane magnet equals the wavelength of the external laser. 
The scheme allows flexibility in the lasing sequence, permitting either the low-energy or high-energy electrons to lase first. In our implementation, the low-energy electrons lase in the first undulator section, resulting in an increased energy separation between the two groups. This choice leads to a modest reduction in the required precision for beam diagnostics and orbit control downstream.
In the first undulator section, high-energy electrons travel off-axis along the undulator and remain fresh slices, while low-energy electrons travel on-axis, generating the first-color mode-locked FEL.
In the second undulator section, the high-energy electrons are redirected onto axis to lase the second-color mode-locked FEL. 
Between the two sections, quadrupoles with transverse offsets are employed as kickers to control the electron bunch orbit. The mode-locked FELs generated in the two undulator sections operate independently, and the time delay between the two colors is adjusted via a magnetic chicane located near the kickers.

To support practical implementation, we outline the diagnostic methods, tuning procedures, and feedback mechanisms that make the proposed scheme experimentally feasible.  
An X-band transverse deflecting structure \cite{craievich2020} combined with a downstream scintillator screen enables sub-femtosecond resolution of the longitudinal phase space. Pulse energy can be measured using a gas monitor detector \cite{sorokin2019x}, while further temporal and spectral characterization can be achieved with a variable-line-spacing  \cite{chuang2017} grating-based spectrometer and a co-axial velocity map imaging spectrometer \cite{li2018}. During commissioning, the modulation laser is initially turned off \cite{duris2021} to simplify beam tuning. Beam position monitors are used to measure the transverse positions of different energy slices, allowing optimization of the kicker settings to align the target slices on-axis in their respective undulators. The effectiveness of these diagnostics and procedures has been experimentally demonstrated \cite{wang24,guo2024}, ensuring that the precision of orbit control required for the proposed scheme is well within the capabilities of modern FEL facilities. In long-term operation, stable performance can be achieved through advanced feedback techniques such as control algorithms based on machine learning \cite{sanchez17} and beam-based orbit correction \cite{sun2024e}.  

In order to make a linear optics analysis of the proposed scheme, we adopt the beam transfer matrix notation of a 6 $\times$ 6 matrix for the beam vector defined by $\vec{e}=\left(x, x^{\prime}, y, y^{\prime}, s, \delta\right)$, where $x$, $y$ and $s$ are respectively the horizontal, vertical and longitudinal coordinates, $x^{\prime} =$ d$x/$d$s$ and $y^{\prime}=$ d$y/$d$s$ are respectively the horizontal and vertical divergences and $\delta$ is the relative energy deviation with respect to the reference particle \cite{chao2023}. 
The transverse structure can be imprinted into either the horizontal direction or the vertical direction. The primary difference is that the magnetic field in $y$ direction decreases much faster than in $x$ direction in a planar undulator. 
For manipulation in $x$ direction, the center of the wiggler and the undulators are on the same horizontal plane. 
Here we leave $(y, y^{\prime})$ out for simplicity, i.e., $(x, x^{\prime}, s, \delta)$ is used in the following and assuming Gaussian distributions for $x, x^{\prime}$ and $\delta$. 

The electron beam is first sent through a dogleg whose transfer matrix is 
\begin{equation}
\boldsymbol{R}_{1} =\begin{pmatrix}
1 & L_{1} & 0 & \eta_1 \\
0 & 1 & 0 & 0 \\
0 & \eta_1 & 1 & \xi_1 \\
0 & 0 & 0 & 1
\end{pmatrix},
\end{equation}
where $L_{1}$ is the length of the first dogleg, $\eta_1$ and $\xi_1$ are the dispersion and the momentum compaction generated in the first dogleg, respectively. After passing through the first dogleg, the electron interacts with the seed laser with the wavelength of $\lambda_s$ in the modulator and gets an energy change $\delta_1= \delta+ \Delta \gamma $sin($2\pi s/\lambda_s$)$/\gamma$, where $\gamma$ and $\Delta \gamma$ are the initial Lorentz factor and its maximum variation after modulation.
 In order to perform the matrix calculations, we approximate the sinusoidal modulation with a linear function $\delta_1=\delta+ hs$, where $h$ varies with $s$.
For electrons at an arbitrary and constant $s$ in the range of $-\lambda_s/2\leqslant s\leqslant\lambda_s/2$,
assuming the reference electrons to be the electrons with zero energy modulation, we have
$h=\Delta \gamma \sin(2\pi s/\lambda_s)/(\gamma s)$. 
Considering electrons in one seed wavelength range, only the electrons near the largest energy modulation positions lase in the undulator sections. For electrons with the largest energy modulation, 
\begin{equation}
 h= 4 \Delta \gamma/(\gamma \lambda_s), s = \pm\lambda_s/4.
\end{equation}Then the corresponding transfer matrix for the modulation part
of electron beam is derived as
\begin{equation}
\boldsymbol{R}_{M} =\begin{pmatrix}
1 & L_{M} & 0 & 0 \\
0 & 1 & 0 & 0 \\
0 & 0 & 1 & \xi_M \\
0 & 0 & h & 1
\end{pmatrix},
\end{equation}
where $ L_{M}$ is the length of the modulator and $\xi_M$ is the momentum compaction generated in the modulator. The momentum compaction of the modulator is usually much smaller than that of the following dispersion section. Thus, $\xi_M$ is ignored in the following calculation.
After the modulator, the energy modulation is converted into temporal-transverse modulation by the downstream dogleg with transfer matrix of 
\begin{equation}
\boldsymbol{R}_{2} =\begin{pmatrix}
1 & L_{2} & 0 & \eta_2 \\
0 & 1 & 0 & 0 \\
0 & \eta_2 & 1 & \xi_2\\
0 & 0 & 0 & 1
\end{pmatrix},
\end{equation}
where $L_{2}$ is the length of the downstream dogleg, $\eta_2$ and $\xi_2$ are respectively, the dispersion and the momentum compaction generated in the downstream dogleg. The transfer matrix for the whole beam line before the undulator is:
\begin{eqnarray}
&&\boldsymbol{R}=\boldsymbol{R}_{2} \cdot \boldsymbol{R}_M \cdot \boldsymbol{R}_{1} \nonumber\\&&=\begin{pmatrix}
1 & L_T+h\eta_2\eta_1 & h\eta_2 & \eta_1+\eta_2+h\eta_2\xi_1\\
0 & 1 & 0 & 0 \\
0 & \eta_1+\eta_2+h\eta_1\xi_2 & 1+h \xi_2 & \xi_1+\xi_2+h\xi_1\xi_2\\
0 & h\eta_1 & h &1+ h\xi_1
\end{pmatrix},
\end{eqnarray}
where $L_T=L_1+L_M+L_2$. In a chicane the two doglegs have opposite polarities, so we have $\eta_2=-\eta_1=\eta$ and $\xi_1=\xi_2=\xi$. Considering the structure of actual device, the relation between $\eta$ and $\xi$ is expressed by $\eta= L\theta$ and $\xi= L\theta^{2}$, where $L=L_1=L_2$ and $\theta$ is the value of the bending angle of the dipoles in each dogleg. The whole transfer matrix after simplification is expressed by:
 \begin{eqnarray}
\boldsymbol{R}=\begin{pmatrix}
1 & L_T-hL^2\theta^2 & hL\theta & hL^2\theta^3\\
0 & 1 & 0 & 0 \\
0 & -hL^2\theta^3& 1+hL\theta^2 & 2L\theta^2+hL^2\theta^4\\
0 & -hL\theta & h &1+hL\theta^2
\end{pmatrix}.
\end{eqnarray}
Subsequently, the transverse position of the electron after the downstream dogleg is written as
\begin{equation}
x_1=x+(L_T-h\eta^2)x^{\prime}+h\eta s+h\xi\eta\delta.
\end{equation}%$h\eta s\approx\eta \Delta \gamma/ \gamma$
For electrons with the largest modulation amplitude,
the term $h\eta s = \eta \Delta \gamma/ \gamma$ indicates a strong coupling between the transverse modulation and the longitudinal energy modulation. The terms $x$, $(L_T-h\eta^2)x^{\prime}$ and $h\xi\eta\delta$ represent the Gaussian background noise from $x$, $x^{\prime}$ and $\delta$,  respectively.
To achieve sufficiently transverse separation of high-energy and low-energy electrons, the condition $ \eta\Delta \gamma/ \gamma \gg (L_T-h\eta^2)\sigma_{x^{\prime}}+h\xi\eta \sigma_{\delta}+\sigma_x$ must be satisfied. 
Here, $\sigma_x$ and $\sigma_{x^{\prime}}$ denote the rms values of $x$ and $x^{\prime}$, respectively, while $\sigma_{\delta}$ corresponds to the rms relative energy spread.
Moreover, a properly optimized $\xi$ can enhance the current density in the lasing part of the beam. For practical parameters, $\sigma_x$ and $\sigma_{x^{\prime}}$ are constrained by the beam emittance, while $\xi$ and $\eta$ are limited by $L$ and $\theta$ of the dogleg.

In the absence of the upstream dogleg, equation (5) implies that the transverse positions of the electrons after the downstream dogleg are given by
\begin{equation}
x_1=x+(L_D+L_M)x^{\prime}+h\eta s+\eta\delta.
\end{equation}
Comparison between equation (7) and equation (8) shows that, since $\eta \gg h\eta \xi $ generally holds, the upstream dogleg effectively avoids the introduction of much larger background noise in the modulation.

\section{Simulations}

Analytical linear calculations give a rough estimation of the optimized condition. Three-dimensional numerical simulations are necessary to show the possible performance of the proposed scheme. 
The following simulations are performed using the same parameters listed in Table~\ref{tab:table1}. In addition, the initial beam sizes are set to $\sigma_x =$ 60 $\upmu$m and $\sigma_{x^{\prime}} =$ 1.3 $\upmu$rad. Each dogleg has a length of 5 m, with a deflection angle of 6 mrad at each dipole. Due to the relatively small $R_{56}$ and mild bending angles of the doglegs and the chicanes, the impact of coherent synchrotron radiation on beam quality is found to be negligible in our configuration.
The electron beam is tracked through the modulator and the doglegs using the code ELEGANT \cite{BorlandM2000} with second-order transport effects taken into account. The FEL lasing processes were simulated using the time-dependent mode of GENESIS \cite{Reiche1999}. These simulations considered realistic transfer matrices of the elements in the proposed scheme. Various nonlinear effects such as coherent synchrotron radiation, incoherent synchrotron radiation, and longitudinal space charge effects are also included in these simulations.

\subsection{Temporal-transverse modulation}
The phase space at the end of the wiggler is identical to that shown in Fig. \ref{fig1}(a) and Fig. \ref{fig1}(c). With a modulation amplitude of 10 MeV, a large value of $\eta$ is not required. In this case, $\eta=$ 0.03 m, which results in a transverse separation of the high-energy electrons relative to the reference particle of  $\eta \Delta \gamma/ \gamma \approx$ 120 $\upmu$m, corresponding to a total separation between the high-energy and low-energy electrons of approximately 240 $\upmu$m.

\begin{figure}[htb]
    \centering
    \includegraphics[width=1\linewidth]{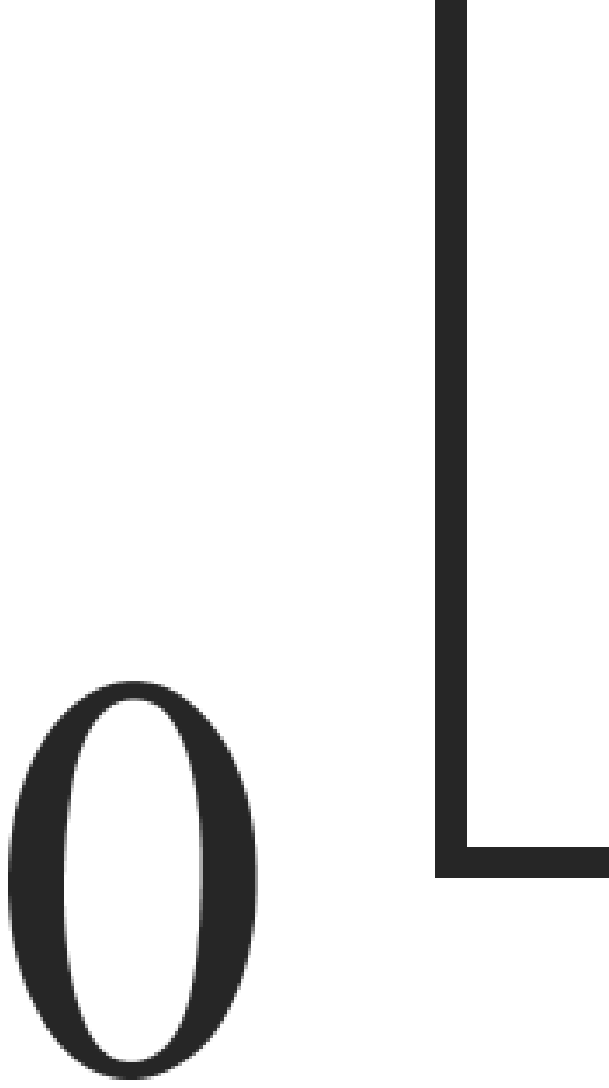}
    \caption{The (a) longitudinal phase space; (b) density distribution and (c) current profile of the electron beam at the entrance of the undulator.}
    \label{fig2}
\end{figure}

To provide sufficient slippage length for the generation of mode-locked FEL pulse trains, the longitudinal length of the electron beam with nearly constant energy is expected to exceed 20 times the seed laser wavelength. In this study, the simulated electron beam length is 24 $\upmu$m.
Since the proposed scheme relies on transforming energy modulation into temporal-transverse modulation, a relatively uniform peak modulation amplitude across the entire  electron bunch is desired. 
Electrons contributing to lasing should experience similar energy modulation to maintain consistent transverse separation and FEL gain. This requires both spatial overlap and temporal synchronization between the laser pulse and the electron bunch to be stable throughout the interaction region.
We note that while this configuration may appear demanding, the required precision is in fact compatible with current FEL infrastructure.
For instance, based on the parameters in Table~\ref{tab:table1}, the scheme tolerates up to 10 fs of timing jitter between the laser and the electron beam, and can operate with a laser intensity jitter below 5\%. Such levels of stability have already been demonstrated at seeded FEL facilities such as FERMI \cite{2021seed}.

Figure~\ref{fig2} shows the phase space of the electron beam at the entrance of the undulator. 
For brevity, 3 cycles are depicted, each with a length of 800 nm. 
As shown in Fig.~\ref{fig2}(a), the sinusoidal energy modulation is overcompressed, resulting in peak current densities at the extrema of the energy distribution. Electrons with Lorentz factors of approximately 4930 $<\gamma<$ 4934 are regarded as the low-energy portion, while those with 4968 $<\gamma<$ 4972 are regarded as the high-energy portion.
In Fig.~\ref{fig2}(b),  
the rms transverse size of both the low-energy electrons and the high-energy electrons is about 60 $\upmu$m, and
the centers of the highest-energy electrons and the lowest-energy electrons are separated by 200 $\upmu$m. 
The on-axis electron beam in each modulation cycle has a transverse extent of approximately 150 nm in the undulator sections, corresponding to a temporal width of about 500 as.
Figure~\ref{fig2}(c) shows that the peak current reaches 5 kA, with the on-axis component contributing approximately 4.2 kA.  
This periodic temporal-transverse modulation of the electron beam satisfies the requirements for subsequent orbit control and two-stage FEL amplification, as will be discussed in the following sections.

\subsection{Orbit control along the undulators}
The first undulator stage consists of 16 modules, with a resonant wavelength of 2 nm. 
Each undulator segment introduces a slippage of 40 nm, while the subsequent chicane contributes an additional slippage of 760 nm per module. 
The second undulator section comprises 22 modules, tuned to a resonant wavelength of 3 nm. In this stage, the slippage per undulator and per chicane is 60 nm and 740 nm, respectively. 
An adjustable chicane is inserted between the two sections to control the time delay of the two pulses,  which is initially set to be 0 in the simulations presented below. 
Two quadrupoles with transverse offset are used as kickers to steer the electron bunch orbit between the two undulator sections. 
The deflection angles of the two kickers are 43.1 $\upmu$rad and $-$65.0 $\upmu$rad, respectively, with a spacing of 1 m between them. 
To ensure FEL saturation, more undulator modules are used in the second stage, mainly due to the slightly larger on-axis orbit oscillation amplitude compared to the first undulator section.
\begin{figure}[htb]
    \centering
    \includegraphics[width=1\linewidth]{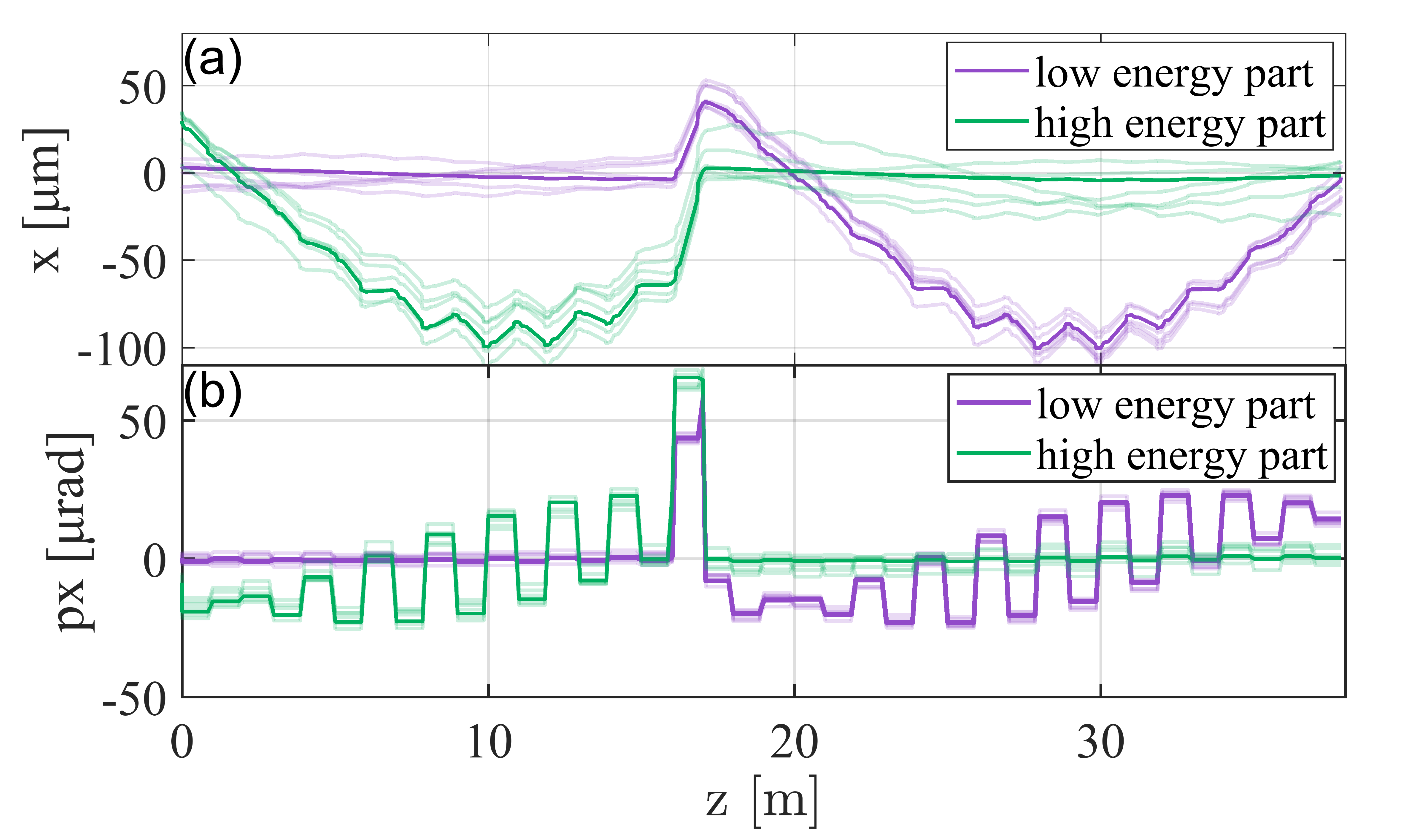}%\{Fig3_21_.eps}
    \caption{Orbit control along the undulators. Average $x$ (a) and average $x^{\prime}$ (b) of the selected electrons along the undulator position $z$. Solid purple and green lines represent the trajectories of the lowest- and highest-energy electrons, respectively; the corresponding light-colored lines show adjacent slices with similar energies.}
    \label{fig3}
\end{figure}

The orbits of the bunch along the undulator sections are shown in Fig.~\ref{fig3}. 
The solid purple and green lines depict the transverse trajectory and angle of the lowest-energy and highest-energy electron slices, respectively, while the lighter curves correspond to slices with energies close to these two extreme-energy groups. In the first undulator section, the oscillation amplitude of the low-energy electrons is approximately 2 $\upmu$m to 10 $\upmu$m, which is  smaller than the electron beam size and the FEL spot size. These electrons are therefore considered to be effectively on-axis, allowing efficient lasing. In contrast, the high-energy electrons exhibit an oscillation amplitude of approximately 100 $\upmu$m, which is larger than the rms transverse size of the on-axis low-energy electrons. Consequently, they are effectively off-axis, resulting in poor spatial overlap and suppression of FEL gain.
After passing through the two kickers, the orbits are reversed: the high-energy electrons are corrected to an oscillation amplitude of 3 $\upmu$m to 20 $\upmu$m, while the low-energy electrons are deflected to about 100 $\upmu$m, becoming off-axis.

These results also demonstrate that the proposed orbit control tolerates moderate energy deviations. Even when the electron energy differs from the nominal value by a few MeV, the corresponding oscillation amplitude remains within 10 $\upmu$m to 20 $\upmu$m. In this range, the beam remains sufficiently on-axis to maintain lasing. 

\subsection{Performance of the proposed scheme}

It has been demonstrated that the evolution of the radiation field in an undulator in the absence of the high-gain FEL interaction may be approximated by the following wave equation \cite{thompson08,kur2011}: 
\begin{equation}
\frac{\partial A\left(\bar{z}, \bar{z}_1\right)}{\partial \bar{z}}+\frac{\partial A\left(\bar{z}, \bar{z}_1\right)}{\partial \bar{z}_1}=b_0\left(\bar{z}_1\right),
\end{equation}
where the scaled units introduced in \cite{bonifacio1989} are used: $A$ is the scaled electric field, $\bar{z}=z/l_g$ is the interaction length along the undulator in units of the one-dimensional gain length $l_g$, $\bar{z}_1\approx (z-\bar{v}_z t)/l_c$ is the  local electron bunch coordinate in units of the cooperation length, $b_0(\bar{z}_1)$ is a small initial electron bunching source term assumed constant with respect to $\bar{z}$. 
For a series of $N$ undulator and chicane modules, 
the solution was given by \cite{thompson08,kur2011}:
\begin{equation}
\tilde{A}(\bar{L}, \bar{\omega})=\tilde{b}_0 \bar{l} \operatorname{sinc}(\bar{\omega} \bar{l} / 2) \frac{e^{i N \bar{\omega} \bar{s}}-1}{e^{i \bar{\omega} \bar{s}}-1} e^{-i \bar{\omega} \bar{l} / 2}.
\end{equation}  
Here $\bar{l}$ is the slippage in one undulator module normalized by $\bar{z}_1$, $\bar{L} = N \bar{l}$ is the scaled interaction length. 
Assuming the slippage due to the chicane in one module is $\bar{\delta}$, the total slippage is expressed by $\bar{s} = \bar{L} +\bar{\delta}$.
And the slippage enhancement factor is defined to be $S_e = \bar{s}/\bar{l}$.
The intensity of the spectrum as a function of  $\omega$ is expressed by:
\begin{eqnarray}
&&I^{(N)}(\bar{\omega}) \sim\left|\tilde{A}^{(N)}(\bar{\omega})\right|^2\nonumber\\&&
=\left(\tilde{b}_0(\bar{\omega}) \bar{l} \operatorname{sinc}(\bar{\omega} \bar{l} / 2)\right)^2 \frac{1-\cos (N \bar{\omega} \bar{s})}{1-\cos (\bar{\omega} \bar{s})}.
\end{eqnarray}
Equation (11) indicates a comb structure in the spectrum, with the number of the modes being approximately $2S_e-1$, and the scaled mode spacing of $\Delta \bar{\omega}=2\pi/\bar{s}$. 
To maximize the number of spectral spikes, it is essential that the slippage within a single undulator remains shorter than the slippage over one gain length.
The total spectrum bandwidth is calculated as $\Delta \omega / \omega_r \approx 1.2 / N_u$. Here $N_u$ is the period number in a single undulator. 
In Ref. \cite{kur2011}, a discrepancy between the analytical prediction and simulation results was reported. The high-gain FEL interaction may lead to a bandwidth narrowing during amplification, reducing the total spectral width by up to 50\%. A similar narrowing behavior is also observed in our following simulation results. 
Taking the peak current to be 4200 A and other parameters listed in Table~\ref{tab:table1}, the  photon energy interval of each mode is calculated to be $\Delta E_{ph}= hc/{\lambda_s}$= 1.553 eV in both the undulator sections. Ignoring the influence of the complex transverse distribution at the current peak, the Pierce parameter in the first undulator section is estimated to be $\rho \approx 3.61 \times10^{-3}$ yielding a gain length of $l_g = \lambda_u/(4\pi\rho)$ = 0.661 m. Assuming $l_{coh} \approx l_c$, the coherence time is $\tau_{coh} \approx \lambda_r/(4\pi c\rho )= $147 as. In the second undulator section, the Pierce parameter is estimated to be $\rho \approx 5.1 \times10^{-3}$, corresponding to a gain length of $l_g $ = 0.471 m and a coherence time of $\tau_{coh} \approx $157 as. 
\begin{figure}[t]
    \centering
    \includegraphics[width=1\linewidth]{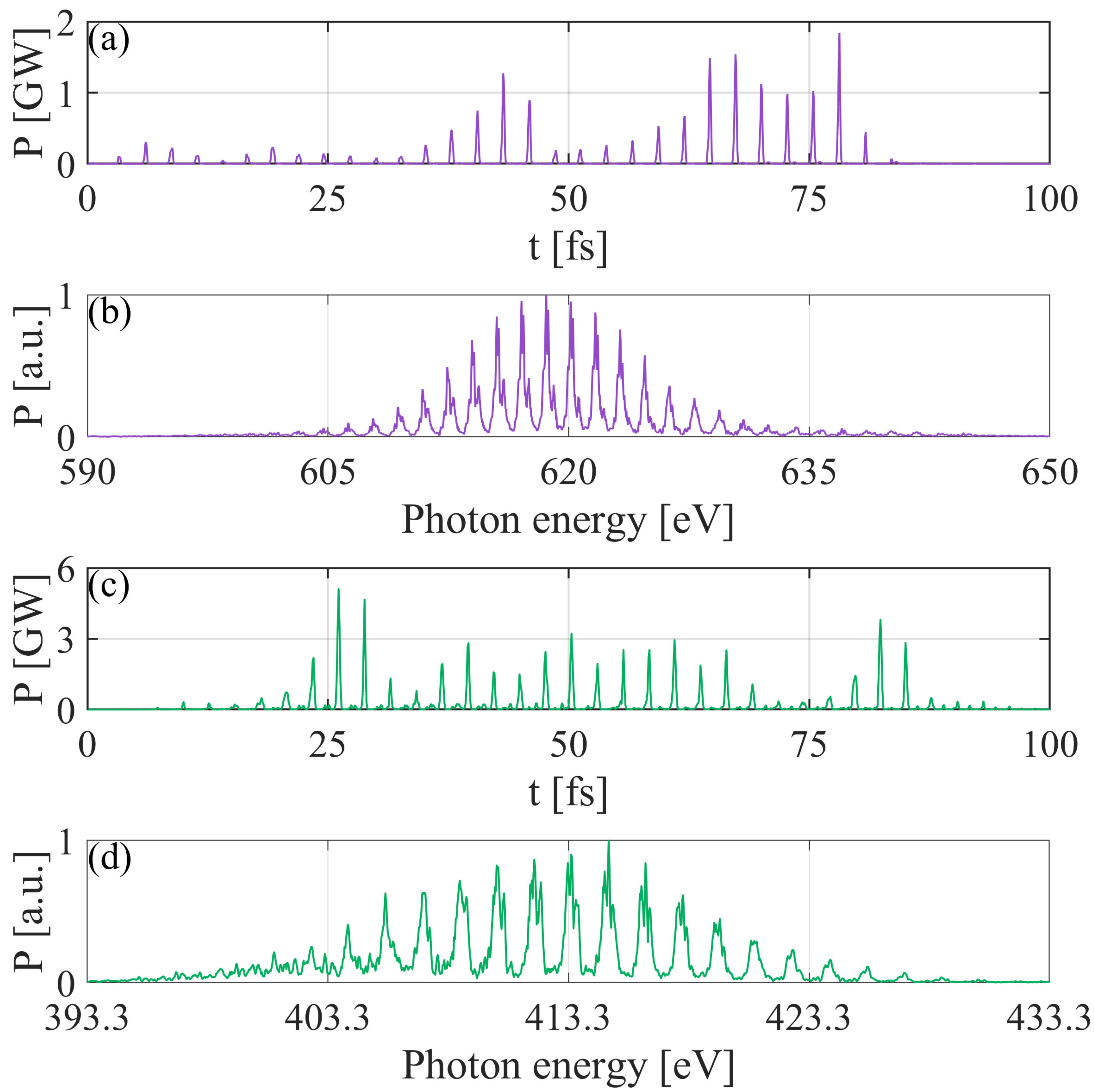}
    \caption{FEL performances of the proposed scheme: (a) radiation power profile and (b) spectrum at the end of the first undulator section; (c) radiation power profile and (d) spectrum at the end of the second undulator section. Bunch head is to the right.}
    \label{fig4}
\end{figure}

The simulation results for a single shot are presented in Fig.~\ref{fig4}, with the time delay introduced by the delay chicane between the two undulator sections set to 0. 
In the first undulator section, Fig.~\ref{fig4}(a) and Fig.~\ref{fig4}(b) show the FEL performance after 13 undulator modules, where the average pulse duration is relatively short, while the pulse energy is also reasonably balanced before reaching saturation. 
Figure ~\ref{fig4}(a) displays a train of pulses evenly spaced by $T_s =$ 2.67 fs, with a total pulse energy of 4.23 $\upmu$J, the average peak power of 0.517 GW and the average pulse duration of $\tau_p =$ 248 as. Due to the slippage induced by the delay chicanes, the total number of the pulses is slightly more than the number of the electron cycles. It is also the reason that the peak power of the pulses near the bunch tail is lower than the average value. 
The peak power of the FEL from the off-axis beam is suppressed below 0.01 GW, 
owing to transverse orbit control discussed in previous sections. 

In Fig.~\ref{fig4}(b), the total pulse train consists of approximately 40 modes, evenly spaced by 1.55 eV. The spectral width (FWHM) of the total pulse train is 2.10\% of the central photon energy, and the average bandwidth of each individual mode is 0.41 eV.
The spectrum is consistent with analytical calculations.
In the second stage, the performances of the FEL at the end of 19 undulators are shown in Fig.~\ref{fig4}(c) and Fig.~\ref{fig4}(d). Figure~\ref{fig4}(c) gives a train of pulses evenly spaced by $T_s =$ 2.67 fs, with the total pulse energy of 19.4 $\upmu$J, the average peak power of 1.63 GW and the average pulse duration of $\tau_p =$ 320 as. 
The average power of the FEL pulses from off-axis reaches about 0.15 GW due to a longer distance of the undulators. 
Meanwhile, as shown in Fig.~\ref{fig4}(d), the background noise in the spectrum is also slightly worse. In this case, the total pulse train consists of approximately 26 modes, evenly spaced by 1.55 eV. The spectral width (FWHM) of the total pulse train is 3.17\%, and the average bandwidth of each individual mode is 0.45 eV.
Comparison between the simulation results and the analytical estimates shows good agreement in the spectral structure. The comb-like spectrum clearly confirms that mode-locking is successfully established in both undulator sections. %Besides, since the complex transverse distribution near the current peak slightly disturbs the gain process, the average pulse duration is slightly longer than the calculation. 

To evaluate the scheme’s robustness against realistic laser jitter, we examine two perturbed scenarios: (i) a fixed timing offset of 10 fs between the modulating laser and the electron beam, and (ii) increasing the laser peak power by 5\% from 40 GW to 42 GW.
In single-shot results, the structured spectrum and bandwidth of individual spectral modes remain stable, as they are governed by the averaged properties of the entire pulse train. The temporal structure, however, is more susceptible to shot-to-shot fluctuations originating from the random self-amplified spontaneous emission (SASE) noises \cite{bermudez2021}. 
To mitigate the influence of SASE noise and extract meaningful comparisons, we perform ten-shot averaging for each condition. As a performance baseline, we first simulate the nominal case using the parameters listed in Table~\ref{tab:table1}. The resulting pulse energy of the 2 nm FEL at 15 m in the first undulator section is 14.2 $\pm$ 2.09 $\upmu$J, and 25.9 $\pm$ 3.21 $\upmu$J for the 3 nm FEL at 19 m in the second undulator section. Under a timing offset of 10 fs, the corresponding pulse energies are 13.5 $\pm$ 1.68 $\upmu$J and 25.6 $\pm$ 3.06 $\upmu$J, while in the increased-power case, the energies are 12.6 $\pm$ 1.63 $\upmu$J and 25.5 $\pm$ 3.30 $\upmu$J, respectively. These results confirm that the proposed scheme remains reasonably robust to experimentally achievable levels of timing and intensity jitter.

\subsection{Influence of the delay chicane}
Assuming a time delay $\Delta t$ introduced by the chicane between the two undulator sections, the corresponding momentum compaction is given by $\xi_d=2 c\Delta t$.
The total momentum compaction from the end of the wiggler to the entrance of the second undulator section is expressed as
 $\xi_T= \xi_2+ \xi_{s1}+\xi_d$, where $\xi_{s1}$ denotes the momentum compaction accumulated in the first undulator section. For an undulator section comprising $N_{u1}$  modules,  $\xi_{s1}= 2N_{u1}\lambda_s$. 
In the present case, $\xi_2 = 180$ $\upmu$m and $\xi_{s1} = 16\lambda_{s} = 12.8$ $\upmu$m, with the total delay of $\xi_2+\xi_{s1}$ corresponding to $\Delta t_0 = $ 321 fs. 
As the electron beam is overcompressed at the end of the first undulator section, the peak current gradually decreases with increasing $\xi_d$. 
When $\xi_d \ll \xi_2+\xi_{s1}$, the delay chicane has negligible impact on the phase space of the beam. 
Unlike the mode-locking scheme \cite{kur2011} exploiting enhanced self-amplified spontaneous emission \cite{zholents2005} where the peak current is enhanced by several times, the proposed scheme results in only a modest current enhancement around electrons with the maximum energy modulation. Therefore, the scheme allows for a wider acceptable range of both $\xi_d$ and $\xi_2$.
\begin{figure}[htb]
    \centering
    \includegraphics[width=1\linewidth]{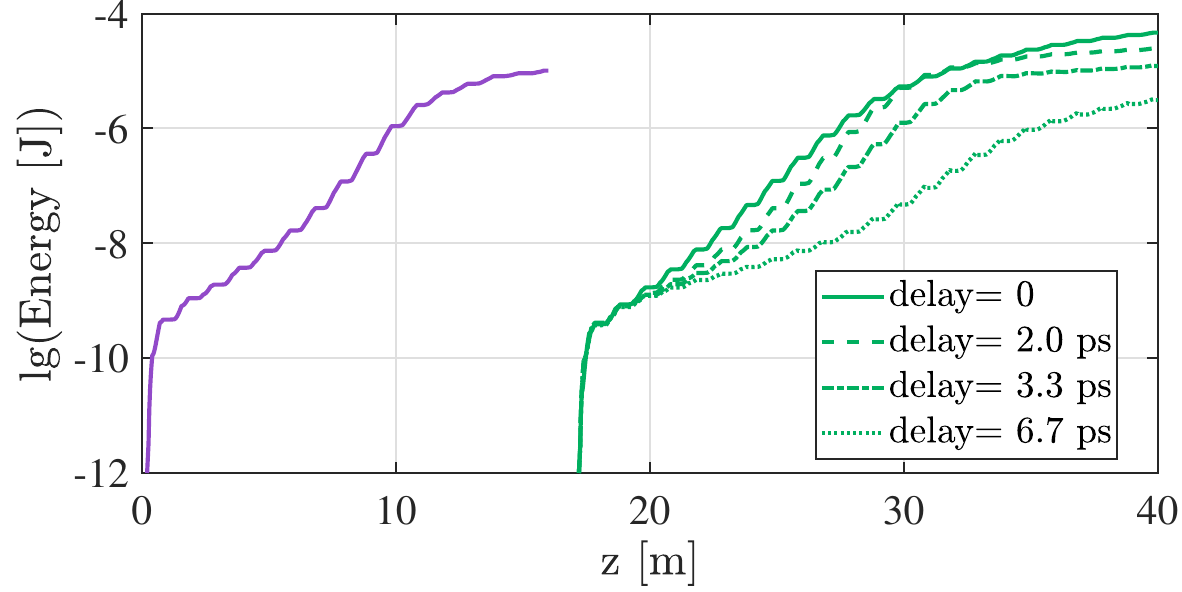}
    \caption{FEL gain curve of the proposed scheme. Purple line represents the gain curve in the first undulator section. Green lines represent the gain curve in the second undulator section under the condition of delay chicane with different values.}
    \label{fig5}
\end{figure}

To assess the impact of the delay chicane, we perform a scan over the delay $\Delta t$ and present the resulting FEL gain curves in Fig.~\ref{fig5}.  
In case of no time delay between the two undulator sections, the solid curves represent the evolution of pulse energy along the undulator: purple for the first undulator section and green for the second. 
Green dotted lines represent the gain curve in the second undulator section with delays of several typical values.
At the end of 19 m in the second undulator section (36 m from the beginning of the first undulator section), the pulse energies are 28.84 $\upmu$J, 19.49 $\upmu$J, 9.77 $\upmu$J and 1.35 $\upmu$J for delay of 0, 2.0 ps, 3.3 ps and 6.7 ps, respectively.
Notably, significant suppression of FEL gain occurs only when $\Delta t$ increases to more than 3.3 ps (about $10 \Delta t_0$ for the simulated parameters). For delays $\Delta t \lessapprox \Delta t_0$, the FEL gain curve closely follows that of the zero-delay case. 

When the delay chicane is set to zero, the two mode-locked pulse trains temporally overlap. 
During the first two undulator modules of each section, both purple and green curves exhibit similar behavior, indicating that the two pulse trains originate independently from random noise in high-energy and low-energy electron slices, respectively. Consequently, their spectral phases remain uncorrelated from shot to shot. However, in scenarios with a small time delay between the two pulse trains, the kicker can be omitted, and subharmonic or harmonic microbunching \cite{guo2024} may be exploited for radiation generation. Under such conditions, the spectral phase between the two pulse trains is also correlated.

In the proposed scheme where an adjustable delay is typically required, as the time delay gradually increases, the pulse trains become increasingly separated, similar to pulling apart two optical combs with precisely defined pulse positions. The minimum achievable time interval between the two pulse trains is determined by the precision of the delay chicane and the timing jitter of the lasing electron slices.
In this standard configuration, if the photon energy of one pulse train is an integer multiple of that of the other, the delay chicane can serve an additional function: it suppresses harmonic microbunching generated in the first undulator section, thereby mitigating potential noise contributions from the off-axis beam that carries residual microbunching. 
 
\subsection{Comparison with the nominal mode-locking scheme}

\begin{figure}[htb]
    \centering
    \includegraphics[width=1\linewidth]{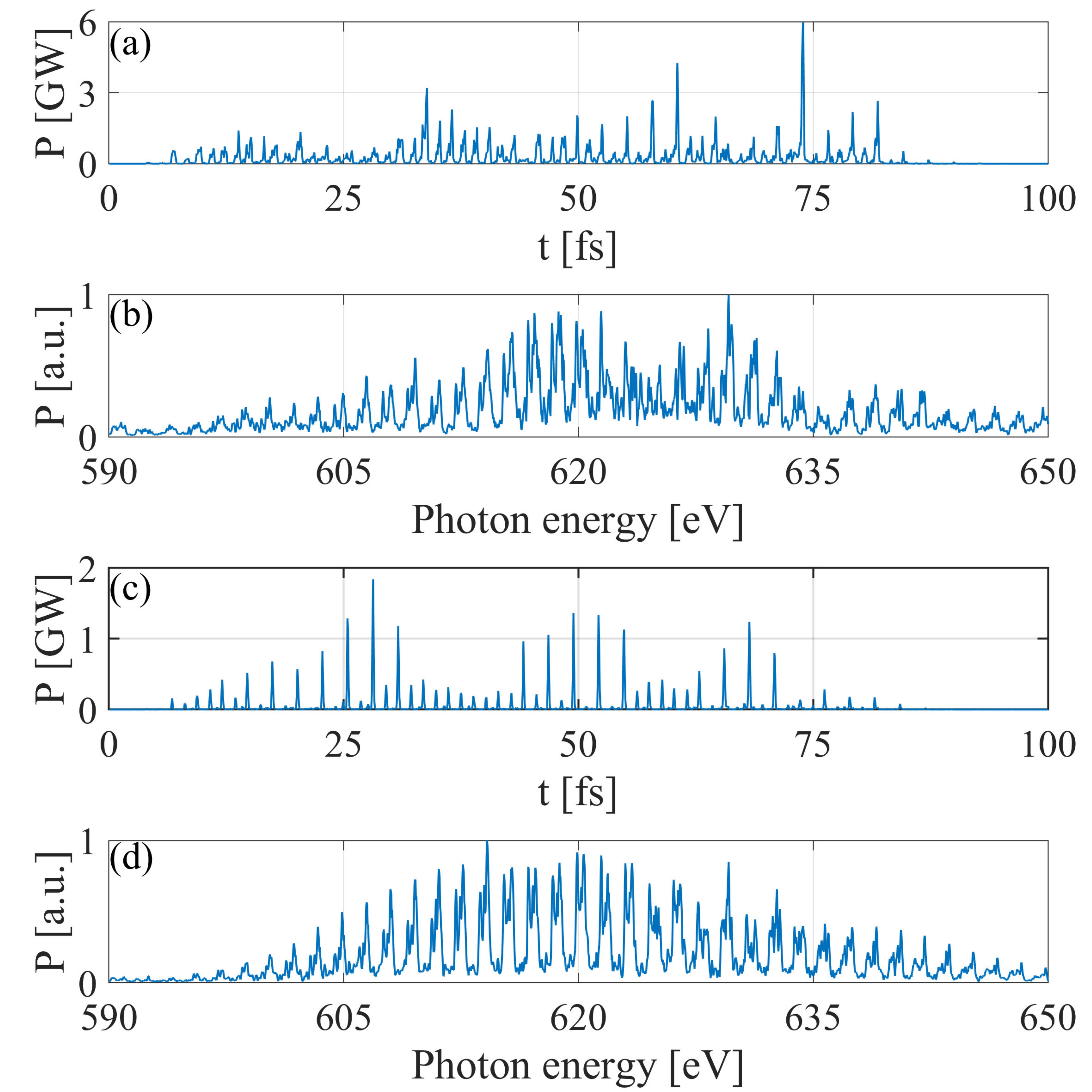}
    \caption{FEL performances of the nominal mode-locked FEL with mostly same parameters: (a) radiation power profile and (b) spectrum at the end of 13 undulators; (c) radiation power profile and (d) spectrum at the end of 9 undulators.}
    \label{fig6}
\end{figure}
To provide a direct comparison between the proposed scheme and the nominal mode-locked FEL, simulations are performed with essentially the same parameters. The phase space and the transverse distribution of the modulated electron beam prior to the undulator are identical to those shown in Fig.~\ref{fig1}(a) and Fig.~\ref{fig1}(c). The longitudinal length of the lowest-energy electrons is approximately 90 nm. The undulator parameters are the same as those used in the first undulator section of the proposed scheme. The Pierce parameter is calculated to be $\rho = 3.23 \times10^{-3}$. 

Figure~\ref{fig6}(a) and Fig.~\ref{fig6}(b) show the FEL power profile and spectrum at the end of 13 undulator modules, corresponding to the same location as Fig.~\ref{fig4}(a) and Fig.~\ref{fig4}(b). 
A notable difference is that although only the low-energy electrons are precisely matched to the lattice, the high-energy electrons still lase.
Similar observations have been reported in Refs. \cite{thompson08,kur2011}, especially for smaller modulation amplitudes.
In this case, the time intervals between pulses are half the wavelength of the seed laser, despite the slippage in the undulator and delay chicane corresponding to one seed laser wavelength.
With larger energy modulation amplitudes, lasing from the high-energy electrons is slightly suppressed but remains significant. 
As shown in Fig.~\ref{fig6}(a), for a single shot the average power and the average pulse duration of all the pulses are 1.22 GW and 474 as, respectively. 
Specifically, the average peak power from the low-energy electrons reaches 1.72 GW, while that from the high-energy electrons is as high as 0.72 GW. 
Since all electrons remain on axis, the average power exceeds that in Fig. ~\ref{fig4}(a).
Under this condition, the electrons within the energy chirps also begin to lase,  increasing the background noise.
Figure~\ref{fig6}(b) presents the spectrum with an overall bandwidth of 3.2\%. 
Although the mode spacing remains at 1.55 eV, the spectrum deviates from a Gaussian shape compared with Fig.~\ref{fig4}(b).  
Increased random noise is observed both across the entire spectrum and within individual modes, a characteristic typically associated with SASE spectra.

Considering that the high background noise at 13 m masks the true characteristics of the nominal case, Figures~\ref{fig6}(c) and ~\ref{fig6}(d) depict the temporal power profile and spectrum at a pre-saturation location of  9 m, where the background noise from chirped electrons is reduced compared to that at 13 m. In Fig.~\ref{fig6}(c), the average peak power and pulse duration are 0.38 GW and 203 as, respectively. The FEL power from the low-energy electrons is 0.62 GW, while that from the high-energy electrons remains higher than in Fig.~\ref{fig4}(a) at 0.14 GW. 
Figure~\ref{fig6}(d) shows a spectrum bandwidth of 4.19\%, with a single mode bandwidth of 0.605 eV. 
Since the delay chicane after each undulator is still 780 nm instead of 380 nm, correlations form every two FEL pulses, and the phases of adjacent pulses are uncorrelated. This results in split peaks within each mode.

These comparisons demonstrate that the proposed scheme effectively suppresses FEL emission from off-axis electrons, resulting in a significantly improved mode-locked FEL performance in both the temporal power profile and spectrum compared to the nominal scheme.

\section{Conclusion}

This study proposes an innovative approach that enables efficient transformation of the laser-induced energy modulation into a periodic temporal-transverse structure, by incorporating a laser modulator within a chicane. 
Compared with existing seeded FEL layouts, we shift the modulator horizontally by several centimeters, and employ two doglegs to provide the dispersion. This structure naturally initiates a fresh-slice based two-color mode-locking process across two undulator sections. As a result, the overall configuration allows independent tuning of the FEL wavelengths and precise control over the temporal separation between the two attosecond pulse trains.

Start-to-end simulations confirm that this scheme can stably generate gigawatt-level attosecond pulse trains with tunable color separation and delay, along with enhanced spectral purity and peak power compared to nominal mode-locked FELs. The required electron beam duration, modulation depth, and timing jitter tolerance are compatible with existing soft X-ray FEL facilities.  
Moreover, applying transverse modulation in the vertical plane reduces the required modulation depth, lowering the seed laser power to below 5 GW.

The proposed wiggler–chicane setup offers a compact and modular interface between external laser modulation and fresh-slice techniques. It is compatible with a wide range of FEL schemes and can be extended to single-color mode-locking, femtosecond two-color pulses driven by few-cycle lasers, or cascaded isolated attosecond pulses. These findings expand the design space for attosecond FEL generation and offer a viable path toward dual-color X-ray spectroscopy.
 
\begin{acknowledgments}
The authors thank Xiaozhe Shen (IASF) for valuable discussions. This work is supported by the Scientific Instrument Developing Project of Chinese Academy of Sciences (Grant No. GJJSTD20220001), the National Natural Science Foundation of China (Grant No. 22288201), DICP (Grant No. DICP I202304), the Strategic Priority Research Program of the Chinese Academy of Sciences: XDB0970000 (subject: XDB0970100) and LiaoNing Revitalization Talents Program XLYC2202030.

\end{acknowledgments}

\bibliography{2CMLFEL}% Produces the bibliography via BibTeX.

\end{document}